\documentclass[amsmath,twocolumn,aps]{revtex4}

\usepackage{longtable}
\usepackage{latexsym}
\usepackage{amssymb}
\usepackage{epsfig}
\usepackage{amsmath}
\usepackage{float}
\usepackage{color}

\newcommand{\be}{\begin{equation}}
\newcommand{\ee}{\end{equation}}
\newcommand{\bea}{\begin{eqnarray}}
\newcommand{\eea}{\end{eqnarray}}

\newcommand{\hatchi}{\hat{\chi}}
\newcommand{\kpe}{k_\perp}
\newcommand{\kpa}{k_\parallel}

\newcommand{\bl}{\mathbf{l}}

\begin{document}

\title{Cross-correlation cosmography with HI intensity mapping}

\author{
A.~Pourtsidou\footnote{E-mail: alkistis.pourtsidou@port.ac.uk},
D.~Bacon\footnote{E-mail: david.bacon@port.ac.uk},
R.~Crittenden\footnote{E-mail: robert.crittenden@port.ac.uk}
}

\affiliation{
Institute of Cosmology \& Gravitation, University of Portsmouth, Dennis Sciama Building, Burnaby Road, Portsmouth, PO1 3FX, United Kingdom
}

\begin{abstract}
The cross-correlation of a foreground density field with two different background convergence fields can be used to measure cosmographic distance ratios and constrain dark energy parameters. We investigate the possibility of performing such measurements using a combination of optical galaxy surveys and HI intensity mapping surveys, with emphasis on the performance of the planned Square Kilometre Array (SKA). Using HI intensity mapping to probe the foreground density tracer field and/or the background source fields has the advantage of excellent redshift resolution and a longer lever arm achieved by using the lensing signal from high redshift background sources. 
Our results show that, for our best SKA-optical configuration of surveys, a constant equation of state for dark energy can be constrained to $\simeq 8\%$ for a sky coverage $f_{\rm sky}=0.5$ and assuming a $\sigma(\Omega_{\rm DE})=0.03$ prior for the dark energy density parameter. We also show that using the CMB as the second source plane is not competitive, even when considering a COrE-like satellite.
\end{abstract}

\maketitle

\section{Introduction}

Cross-correlation cosmography is a method that uses the gravitational lensing of galaxies in different source (background) redshifts in correlation with a given lens (foreground) population in order to constrain dark energy \cite{Jain:2003tba,Bernstein:2003es,Zhang:2003ii,Taylor:2006aw}. If the foreground distribution is sufficiently narrow in redshift, the ratio of the foreground galaxy density ($\delta_g$) -- background lensing convergence ($\kappa$) cross-correlation from different background bins is a purely geometrical quantity which depends only on the source distribution kernels. In the case that the background distributions are also narrow in redshift, this quantity simplifies to a ratio of comoving radial distances. 

The distance ratio depends on the background cosmology, and in particular on the amount of dark energy and its equation of state;  it is independent of the large scale structure details (power spectrum, bias) and the angular scale. Therefore, measurements at different angular scales - even those in the non-linear regime - can be used to constrain dark energy. Of course, removing the dependence on the growth rate and the associated information it carries about dark energy means that the derived constraints will be weaker than the ones derived via the usual methods which depend on both the geometry and the growth, such as lensing tomography \cite{Hu:2002rm,Abazajian:2002ck,Takada:2003hy,Takada:2003ef}. However the geometric method directly probes the evolution of the Universe via the redshift-distance relation and it is therefore a robust  consistency check. Furthermore, isolating geometry from structure formation is very useful for General Relativity tests, since modified gravity models can mimic the expansion history of the Universe but have different clustering properties \cite{Clifton:2011jh}.

In this work we investigate these correlations using a combination of HI intensity mapping surveys and galaxy surveys. Intensity mapping is an innovative technique able to map the large-scale structure of the Universe in 3D (see e.g. \cite{Peterson:2009ka}). It uses HI as a tracer of the dark matter density field by measuring the intensity of the redshifted 21cm line over the sky in a range of redshifts without detecting individual galaxies, treating the 21cm sky as a diffuse background, similar to the CMB. In \cite{Santos:2015gra} the potential of the planned Square Kilometre Array (SKA) \cite{ska} to deliver HI intensity mapping maps over a broad range of frequencies and a substantial fraction of the sky was investigated. Using HI intensity mapping to probe the foreground density tracer field and/or the background source fields has the advantage of excellent redshift resolution and a longer lever arm achieved by using precise measurements of the lensing signal from high redshift background sources. 

The paper is organised as follows: In Section~\ref{sec:formalism} we describe the general formalism we will use. We form the tracer density-convergence power spectrum correlating the density from a foreground bin ($f$) and the convergence from a background bin ($b$). We then derive the distance ratio, describe how it depends on dark energy and formulate the corresponding Fisher matrix for the derivation of dark energy constraints. We then move on to the results, Section~\ref{sec:results}. We first consider the usual case of a galaxy survey for both the foreground lenses and the background sources, employing Large Synoptic Survey Telescope (LSST)-like parameters \cite{Ivezic:2008fe}. Then we consider 
combinations of optical galaxy surveys (LSST) with HI intensity mapping surveys performed with SKA-like instruments for measuring the foreground density fields and the background convergence fields. Finally, we show how an even longer lever arm could be achieved by considering 21cm lensing from the Epoch of Reionization, or CMB lensing with a future satellite mission. For all cases considered we derive constraints on the constant equation of state parameter $w_0$ and the dark energy abundance $\Omega_{\rm DE}$. We discuss our results and conclude in Section~\ref{sec:conclusions}.

\section{Formalism}
\label{sec:formalism}

We are interested in the correlation of the foreground ($f$) density tracer field $\delta_{\rm tr}$ with the background ($b$) convergence $\kappa$ field.  We will assume linear biasing, so that $\delta_{\rm tr} = b_{\rm tr} \delta$, where $\delta$ is the underlying matter density field and $b_{\rm tr}$ is the tracer's bias --- in this work, we use `${\rm tr} = g$' for galaxy surveys and `${\rm tr} = {\rm HI}$' for intensity mapping surveys. Using the Limber approximation \cite{Limber:1954zz} the angular cross-power spectrum $C_{\delta_{\rm tr} \kappa}$ is given by
\begin{align} \nonumber
C_{\delta_{\rm tr} \kappa}(\ell; \, f,b)&=\frac{3\Omega_{\rm m}H^2_0}{2c^2}\int \frac{d\chi_f}{a(\chi_f)}W_f(\chi_f)
\int d\chi_b W_b(\chi_b) \\ 
& \times \frac{\chi_b-\chi_f}{\chi_b \chi_f} b_{\rm tr}(\chi_f) P_{\rm \delta \delta}\left(\frac{\ell}{\chi_f},\chi_f\right),
\label{eq:Cdeltakappa}
\end{align} where $\chi$ is the comoving distance and $P_{\delta\delta}$ is the three-dimensional dark matter power spectrum. We have assumed that the foreground galaxy distribution $W_f$ and the background galaxy distribution $W_b$ do not overlap, so that $\chi_f < \chi_b$ always.This assumption is important, as a possible contamination of background sources in front of foreground density tracers can be a problem; the foreground populations have to be completely in front of the background populations for Eq.~(\ref{eq:Cdeltakappa}) to be exact. A comprehensive analysis of this effect for the case where both the foreground and background populations are probed by galaxy surveys is presented in \cite{Zhang:2003ii}. Of course, one can avoid this systematic by selecting the foreground and background populations such that they are clearly separated in redshift. Using HI intensity mapping is advantageous, because it provides very precise redshift information which is not the case when using photometric galaxy surveys. 

If the foreground lens slice is narrow enough in redshift ($\Delta z \simeq 0.1$ is sufficient), we can approximate the foreground redshift distribution as a delta function at a distance $\hat{\chi}_f$, $W_f(\chi_f)=\delta^{\rm D}(\chi_f-\hatchi_f)$. We then find
\begin{align} \nonumber
C_{\delta_{\rm tr} \kappa}(\ell; \, f,b)&=\frac{3\Omega_{\rm m}H^2_0}{2c^2}\left(\frac{b_{\rm tr}(\hatchi_f) P_{\rm \delta \delta}\left(\frac{\ell}{\hatchi_f},\hatchi_f\right)}{a(\hatchi_f)\hatchi_f}\right) \\
&\times
\int d\chi_b W_b(\chi_b) \times \frac{\chi_b-\hatchi_f}{\chi_b } .
\end{align}
By taking the ratio of $C_{\delta_{\rm tr} \kappa}$ measured using two different background distributions $b_1,b_2$ but the same foreground $f$, we see that the details of the large scale structure and the bias cancel out and all that remains is the geometrical ratio of the source distribution kernels. In the limit that the two background sources also have a delta-function distribution (at $\hatchi_{b_1},\hatchi_{b_2}$, respectively) the ratio becomes
\be
 R(z_f,z_{b_1},z_{b_2}) \equiv \frac{C_{\delta_{\rm tr} \kappa}(\ell; \, f,b_1)}{C_{\delta_{\rm tr} \kappa}(\ell; \, f,b_2)} = \frac{(\hatchi_{b_1}-\hatchi_f)/\hatchi_{b_1}}{(\hatchi_{b_2}-\hatchi_f)/\hatchi_{b_2}} .
\label{eq:Ratio}
\ee Note that the $\ell$-dependence (angular scale dependence) also cancels out.
Assuming a flat Universe, the right-hand-side depends only on the dark energy parameters $(\Omega_{\rm DE},w_0,w_a)$, where we have used the parametrisation $w=w_0+w_a(1-a)$ \cite{Linderparam} for the dark energy equation of state.

The ratio $R$ defined in Eq.~(\ref{eq:Ratio}) is observable and depends only on the comoving distances to the foreground and background populations. 
In a flat Universe, the comoving distance is given by
\begin{align} \nonumber
\chi(z)&=\frac{c}{H_0}\int_0^z dz' [(1-\Omega_{\rm DE})(1+z')^3\\
&+\Omega_{\rm DE}(1+z')^{3(1+w_0+w_a)} {\rm e}^{-3w_az'/(1+z')}]^{-1/2}.
\end{align}
We shall now derive dark energy constraints using the Fisher matrix
\be
 F_{pq}=\frac{1}{2}\sum_f\sum^{\ell_{\rm max}}_{\ell_{\rm min}}  \Delta \ell  f_{\rm sky}(2\ell+1) \frac{\partial R^f/ \partial p}{R^f}\left[\frac{1}{\sigma_f^2(\ell)}\right]\frac{\partial R^f / \partial q}{R^f}, 
\ee
where $R^f \equiv R(z_f,z_{b_1},z_{b_2})$ is the ratio corresponding to foreground bin $f$, $(p,q)\equiv(\Omega_{\rm {\rm DE}},w_0,w_a)$ and 
$\sigma_f^2(\ell)$ is the fractional variance of $R^f$ given by \cite{Zhang:2003ii}
\begin{align} \nonumber
\sigma_f^2(\ell) &= [C_{\rm \delta_{\rm tr} \kappa}(\ell,f,b_1)]^{-2}[C_{\delta_{\rm tr}\delta_{\rm tr}}(\ell,f)+{\cal N}_f(\ell)]
[C_{\kappa \kappa}(\ell,b_1) \\ \nonumber
&+{\cal N}_{b_1}(\ell)] +[C_{\rm \delta_{\rm tr} \kappa}(\ell,f,b_2)]^{-2}[C_{\delta_{\rm tr}\delta_{\rm tr}}(\ell,f)+{\cal N}_f(\ell))] \\ \nonumber
&[C_{\kappa \kappa}(\ell,b_2)+{\cal N}_{b_2}(\ell)]
-2C_{\rm \delta_{\rm tr} \kappa}(\ell,f,b_1)^{-1}C_{\rm \delta_{\rm tr} \kappa}(\ell,f,b_2)^{-1} \\ 
&[C_{\delta_{\rm tr}\delta_{\rm tr}}(\ell,f)+{\cal N}_f(\ell)]C_{\kappa \kappa}(\ell,b_1,b_2).
\label{eq:varR}
\end{align}
Here, ${\cal N}_f(\ell)$ is the noise in the foreground density power spectrum $C_{\delta_{\rm tr}\delta_{\rm tr}}(\ell,f)$ measurement, and ${\cal N}_{b_i}(\ell)$ is the noise in the convergence power spectrum $C_{\kappa \kappa}(\ell,b_i)$ measurement. The exact form of these noise terms depend on the kind of survey we are using (see next Section). Finally, $f_{\rm sky}$ is the fraction of the sky our chosen surveys map, and $\Delta \ell$ the binning in $\ell$-space.

The above equation is the Fourier space equivalent of the variance used in \cite{Jain:2003tba} where halo-shear correlations were considered, but in that work only the shot variance contribution was used, i.e. the approximation
\begin{align}\nonumber
\sigma_f^2(\ell) \simeq &[C_{\rm \delta_{\rm tr} \kappa}(\ell,f,b_1)]^{-2}{\cal N}_f(\ell){\cal N}_{b_1}(\ell)+
[C_{\rm \delta_{\rm tr} \kappa}(\ell,f,b_2)]^{-2} \\
&{\cal N}_f(\ell){\cal N}_{b_2}(\ell) 
\label{eq:varRJT03}
\end{align} 
was made  \cite{Zhang:2003ii}. Throughout this work we use the full noise expression of Eq.~(\ref{eq:varR}); however, we will ignore the (weak) covariance between different foreground bins (see \cite{Zhang:2003ii} for details).

\begin{figure}[H]
\centering
\includegraphics[scale=0.65]{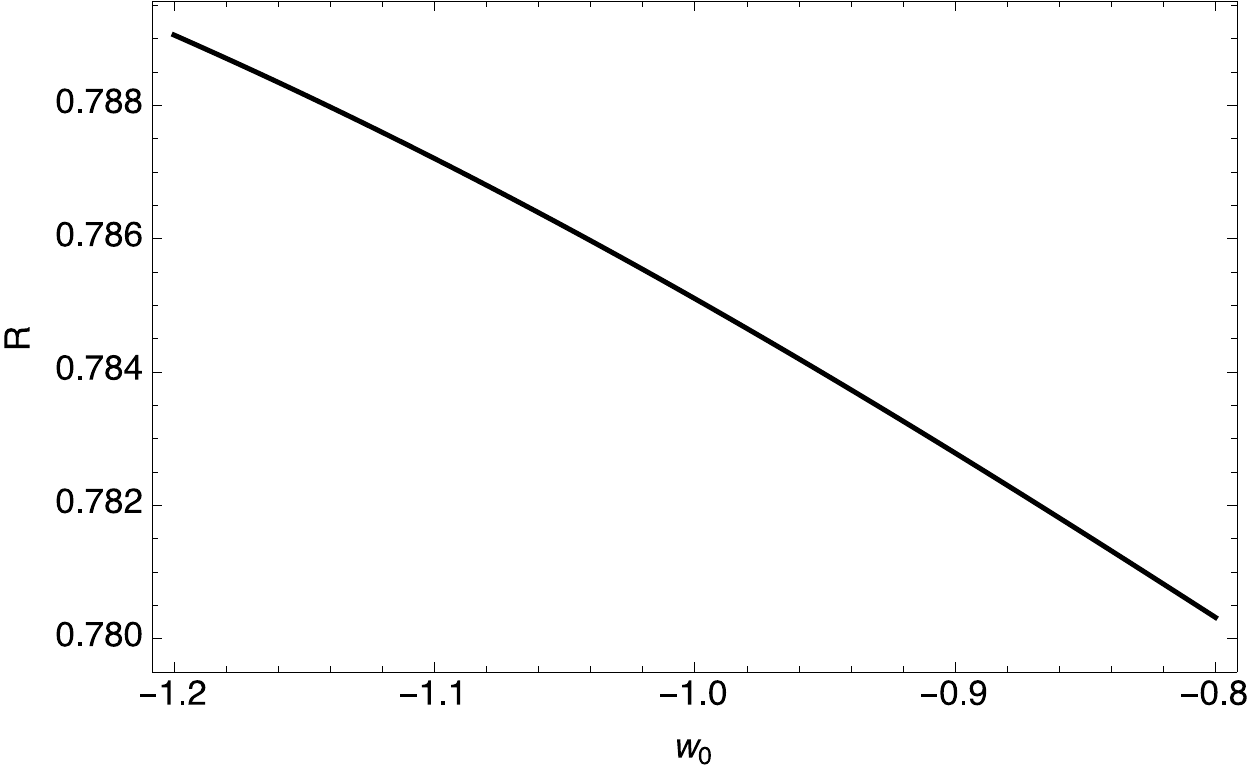}
\caption{The ratio R as a function of $w_0$ for $z_f=0.5$, $z_{b_1}=1.3$ and $z_{b_2}=2.4$, assuming $w_a=0$ and $\Omega_{\rm DE}=0.73$. The dependence on the equation of state is very weak ($R \propto |w_0|^{0.028}$).}
\label{fig:Ratio}
\end{figure}

Before we derive our results using different combinations of intensity mapping and galaxy surveys, it is useful to demonstrate the dependence of the ratio $R$ on the dark energy equation of state. In Figure~\ref{fig:Ratio} we show the dependence of $R$ given by Eq.~(\ref{eq:Ratio}) with $z_f=0.5$, $z_{b_1}=1.3$ and $z_{b_2}=2.4$  on the dark energy parameter $w_0$ (assuming $w_a=0$) with $\Omega_{\rm DE}=0.73$. We immediately see that the dependence is very weak, $R \propto |w_0|^{0.028}$, which means that in order to have good constraints on $w_0$ we would need a very accurate measurement of $R$.

\section{Results}
\label{sec:results}

In the following we will investigate a number of different survey combinations to perform 
cross-correlation cosmography studies. Initially we consider a galaxy survey with LSST-like parameters for both the foreground lenses and the background sources populations.
Then we consider intensity mapping surveys using SKA-like instruments for measuring the foreground density field or the background convergence field, and compare them to the LSST-only results. We also consider the case where our second ($b_2$) background sources are at the Epoch of Reionization. Finally, we use the CMB as the second source plane and a COrE-like satellite \cite{core}. 

When using Eq.~(\ref{eq:Ratio}) for our Fisher matrix calculations we apply the delta function approximation for the distribution of the foreground populations, as they are thin enough in redshift ($\Delta z \leq 0.1$) in all cases. In the case where the background sources are probed with an optical galaxy survey, we use the known redshift distribution of these galaxies. In the 21cm lensing case using the intensity mapping method, the HI galaxies distribution is unknown. However, our chosen bands are narrow enough so that the delta function approximation is valid (i.e. angular distances vary very little within each band). The delta function approximation is also valid when CMB is used as the second source plane.

The survey combinations we considered together with the best obtained $1\sigma$ uncertainties on $w_0$ for each case are shown in Table~\ref{tab:resultssumm}. 

\begin{table}[H]
\centering
\begin{tabular}{|c||c|c|c|c|}
\hline
 Section/Surveys  & $f$    &$b_1$&$b_2$ &$\delta w_0$\\
\hline
\hline
\ref{sec:results}.A/B & LSST & LSST   &  LSST  & 0.12/0.10
\\
\hline
\ref{sec:results}.B & SKA-like & LSST   &  LSST  & 0.08
\\
\hline
\ref{sec:results}.C & LSST & LSST   &  SKA-like  & 0.08
\\
\hline
\ref{sec:results}.D & LSST & LSST   & SKA-like$|_{\rm EoR}$  & 0.11
\\
\hline
\ref{sec:results}.E & LSST & LSST   &  COrE-like$|_{\rm cmb}$  & 0.22
\\
\hline
\end{tabular}
\caption{The survey combinations considered in this work. For each combination the best obtained $1\sigma$ marginalised error on the constant dark energy equation of state $w_0$ is shown.}
\label{tab:resultssumm}
\end{table}

\subsection{Results using a LSST-like survey for the foreground lenses ($f$) and background sources ($b_1,b_2$)}

In the case of galaxy surveys we have $\delta_{\rm tr} \rightarrow \delta_g$ and, consequently, $C_{\delta_g\kappa}$ for the cross-correlation power spectrum, with $\delta_g = b_g \delta$ and $b_g$ the galaxy bias. The noise terms are the shot noise contributions given by
\be
{\cal N}_f(\ell) = \frac{1}{\bar{n}^g_f}\,; \;\;\;  {\cal N}_{b_i}(\ell) = \frac{\sigma^2_\kappa}{\bar{n}^B_i},
\ee
where $\bar{n}^g_f$ is the number density of foreground galaxies in bin $f$, $\sigma^2_\kappa$ is the shape noise of each background galaxy and $\bar{n}^B_i$ ($i=1,2$) is the number density of background galaxies in source bin $i$. 

\begin{figure}[H]
\centering
\includegraphics[scale=0.55]{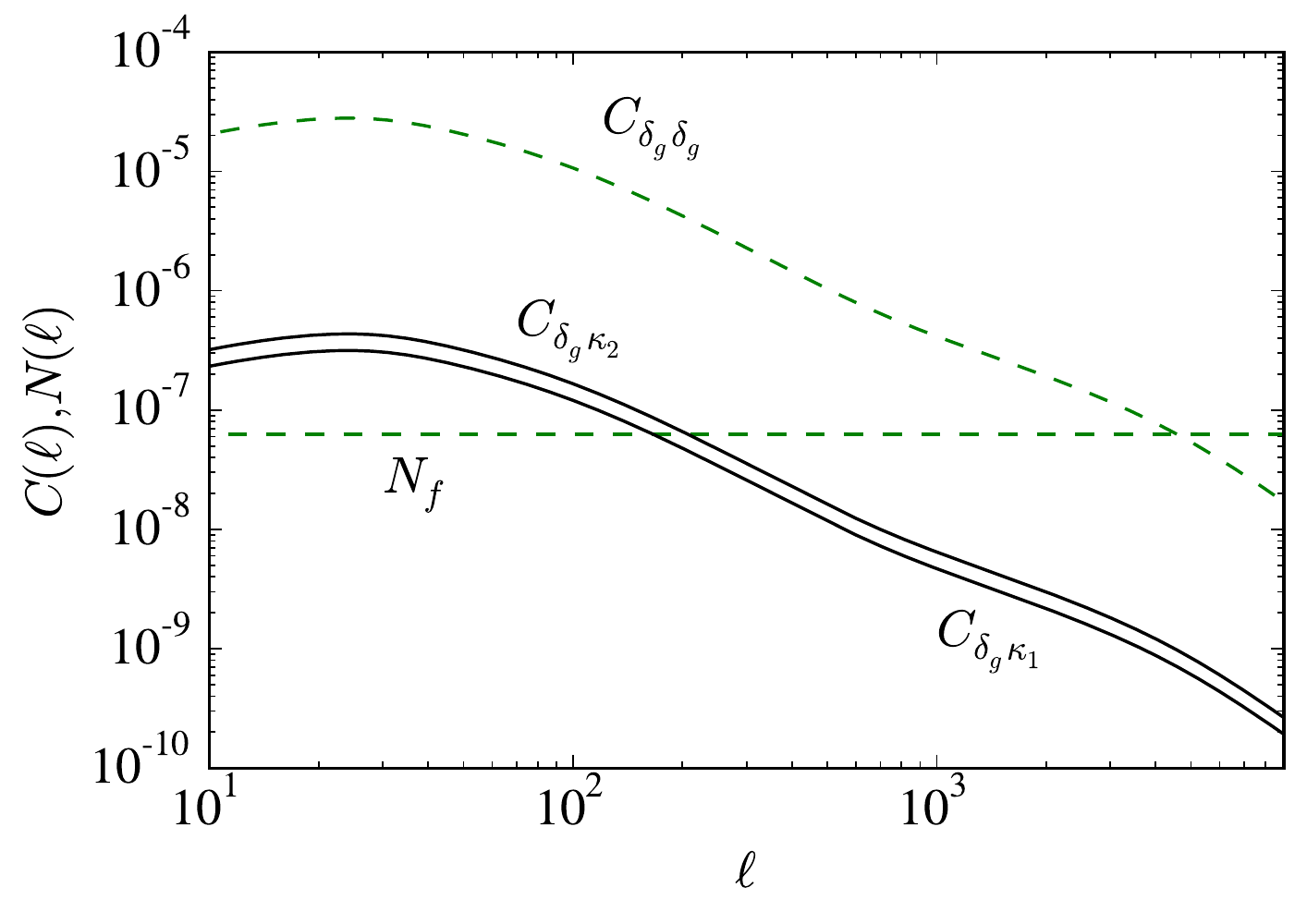}
\includegraphics[scale=0.55]{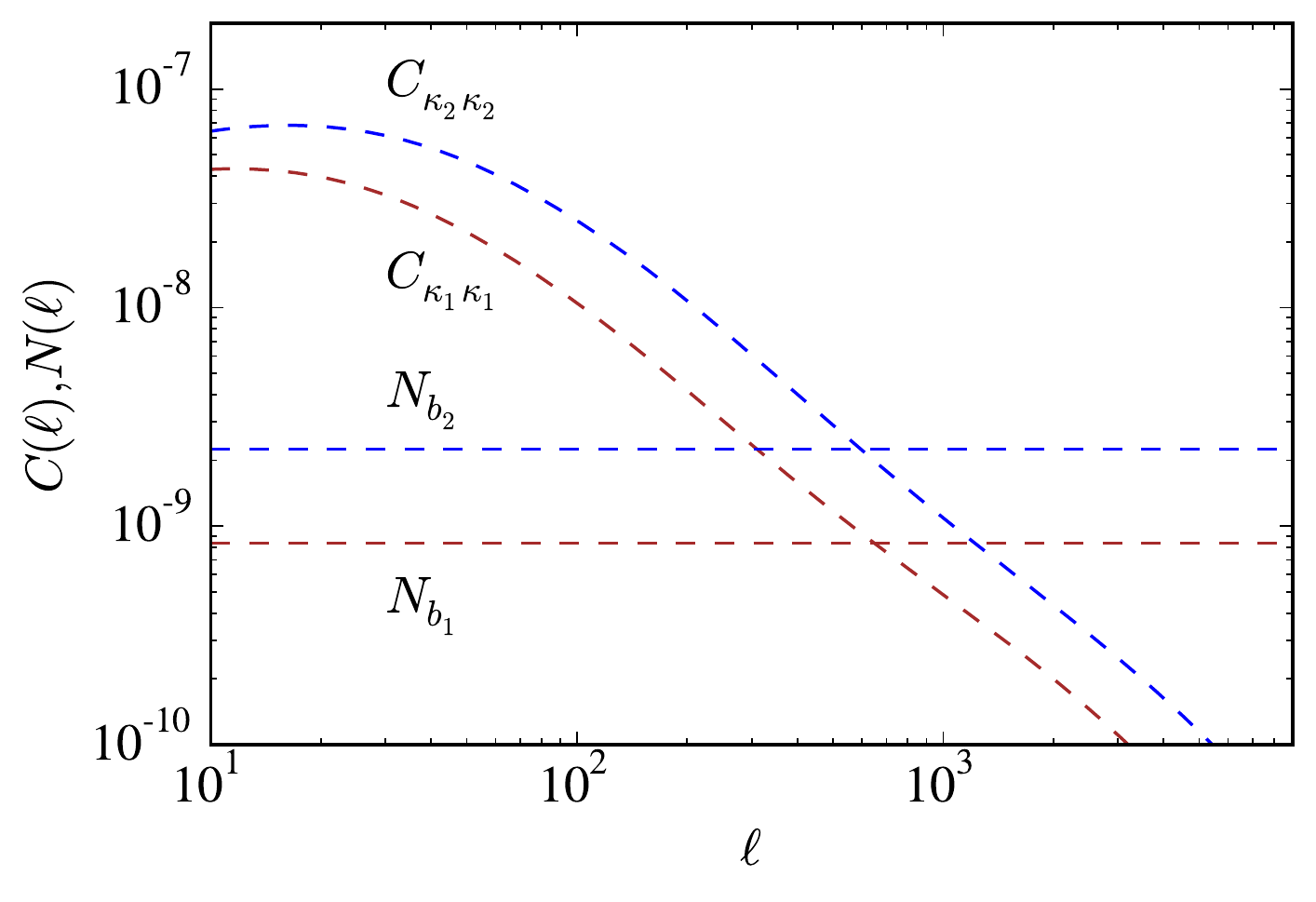}
\caption{Various power spectra that contribute to the error in the measurement of $R^f$, using the $z_f=0.5$ foreground (lens) bin and two background (source) bins with $z_{b_1}=1.3$ and $z_{b_2}=2.4$.}
\label{fig:CgCk}
\end{figure}

We consider an LSST-like galaxy survey  \cite{Ivezic:2008fe} with redshift range $0<z<3$, galaxy number density $40 \, {\rm arcmin}^{-2}$, $\sigma_\kappa = 0.3$, covering half of the sky $f_{\rm sky} = 0.5$.  The redshift distribution of galaxies is modelled by
\be
\frac{dn}{dz} \propto z^2 {\rm e}^{-(z/z_0)^{1.5}}
\ee with $z_0=1$. 
We will consider $10$ foreground top-hat redshift slices from $0.1<z_f<1$ (i.e. $\Delta z=0.1$) and two source (background) populations 
between $z=1.1$ and $1.5$, and between $z=2.2$ and $2.6$ - note that the first background source bin ($b_1$) will always be chosen to be close to the foreground population, as this leads to higher sensitivity of the ratio to the dark energy parameters, hence better constraints. Various auto, cross and noise power spectra that enter the calculation of the error in $R^f$ given by Eq.~(\ref{eq:varR}) are shown in Figure~\ref{fig:CgCk}, using the  $z_f=0.5$ foreground (lens) bin and our two background (source) bins with $z_{b_1}=1.3$ and $z_{b_2}=2.4$. In the Fisher matrix we take bins of $\Delta \ell =30$ starting from  $\ell_{\rm min}=45$ and going up to $\ell_{\rm max}=3000$. To model the various power spectra that enter the error calculation we use the non-linear fitting formula by Smith et al. \cite{Smith:2002dz} and we set the galaxy bias $b_g=1$.

\begin{figure}[H]
\centering
\includegraphics[scale=0.65]{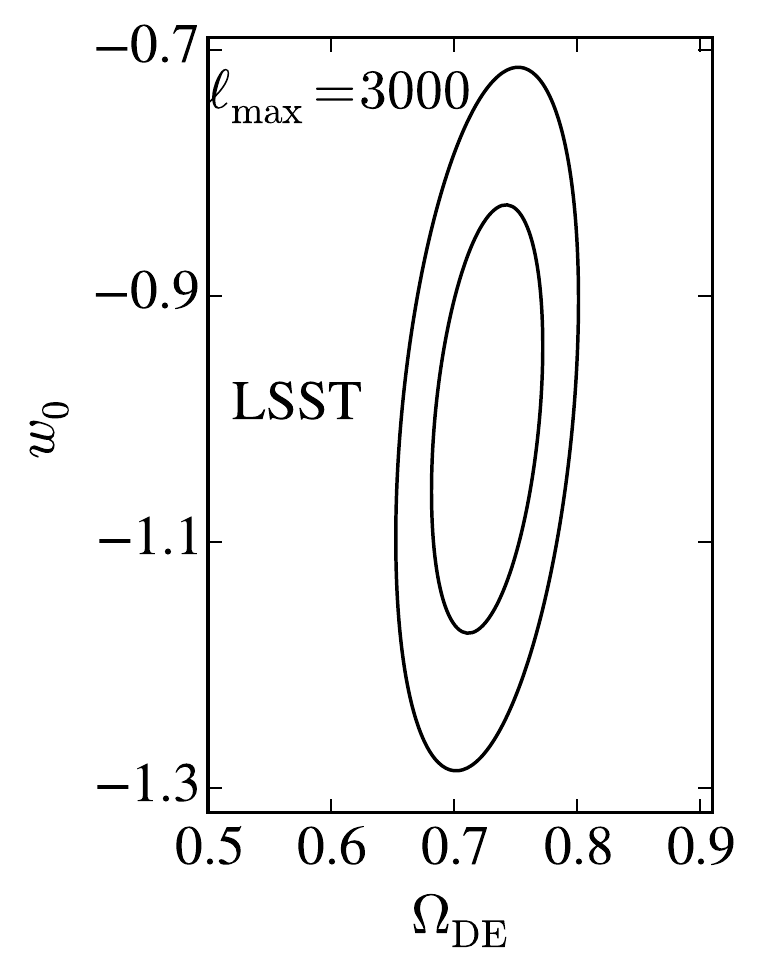}
\caption{Dark energy $1\sigma$ and $2\sigma$ constraints on $w_0$ and $\Omega_{\rm DE}$ fixing $w_a=0$ and employing a prior on $\Omega_{\rm DE}$ corresponding to $\sigma(\Omega_{\rm DE})=0.03$. The fiducial model has $w=-1$ and $\Omega_{\rm DE}=0.73$. We have used an LSST-like galaxy survey with parameters described in the main text.}
\label{fig:LSSTconstr}
\end{figure}

Following previous works (see, for example, \cite{Jain:2003tba,Zhang:2003ii}), we will present constraints on a constant equation of state $w=w_0$ employing a prior $\sigma(\Omega_{\rm DE})=0.03$ for the dark energy density parameter. For the LSST-like survey we find 
\be \nonumber
\delta w_0 \simeq 0.12 
\ee for the marginalised $1\sigma$ $w_0$ uncertainty, while the joint ($\Omega_{\rm DE},w_0$) $1\sigma$ and $2\sigma$ contours are shown in Figure~\ref{fig:LSSTconstr}. Note that if we use the approximation for the squared fractional variance of $R$ used in  \cite{Jain:2003tba}, i.e. Eq.~(\ref{eq:varRJT03}), we find a much smaller $\delta w \simeq 0.04$. This is in qualitative agreement with the findings in \cite{Zhang:2003ii} and shows that, even if there is partial cancellation among the extra terms in the full formula presented in Eq.~(\ref{eq:varR}), they do not cancel exactly and should be taken into account. 

As expected, the constraints we found are weaker than those from methods using the information encoded in the growth function. But the advantage of the geometrical method is that it carries less theoretical assumptions and can be used as a consistency check against methods that use \emph{both} geometry and growth to constrain dark energy. For a detailed discussion on the growth-geometry constraints and issues like photometric redshift accuracy and the use of a more general source scaling of lensing signals which works for both galaxy-shear and shear-shear correlations, see \cite{Zhang:2003ii}. However, note that the aforementioned scaling relation is disrupted by the fact that we can only measure reduced shear and not shear, which introduces a scale dependence on the ratio making it somewhat sensitive to the matter power spectrum \cite{Shapiro:2008yk}.

We will now move on to the interesting possibility of using intensity mapping surveys in combination with galaxy surveys to perform cosmographic studies. We concentrate on the combination of LSST and SKA-like intensity mapping surveys. Note that possible synergies between the LSST and SKA for cosmological probes like galaxy clustering, weak lensing and strong lensing have been recently discussed in \cite{Bacon:2015dqe}.

\subsection{Results using a SKA-Mid-like Intensity Mapping survey for the foreground lenses ($f$) and a LSST-like survey for the background sources ($b_1,b_2$)}

In the case of intensity mapping (IM) surveys we have $\delta_{\rm tr}\rightarrow \delta_{\rm IM}$ 
with $\delta_{\rm IM} = \bar{T} \delta_{\rm HI}= \bar{T} b_{\rm HI}\delta$, where $\bar{T}$ is the mean brightness temperature and $b_{\rm HI}$ the HI bias. 
In a recent publication \cite{Bull:2014rha} the majority of current and planned HI intensity mapping experiments at redshifts $z < 4$ were analysed. Intensity mapping at radio frequencies has a number of attractive features, a very important one being the fact that we can automatically measure redshifts with extremely high precision, which circumvents the photo-$z$ uncertainties that can be a serious source of systematic error in cross-correlation cosmography.

Focusing on the SKA-Mid instrument, we can consider an intensity mapping survey in which the instrument operates in either  ``single-dish autocorrelation" or  ``interferometer" mode \cite{Bull:2014rha}. We will use the latter for our analysis since the single-dish mode does not probe the small scales due to the limited angular resolution which is fixed by the size of the dish. On the other hand, the angular resolution of the interferometer is fixed by the array's maximum baseline, hence it can probe very small angular scales (equivalently, high multipoles $\ell$). 

In this case, the noise contribution is dominated by the thermal noise of the instrument which is calculated using the formula (note we use the uniform distribution approximation) \cite{Zaldarriaga:2003du,Morales05,McQuinn:2005hk}
\begin{equation}
\label{eq:CellN}
C^{\rm N}_\ell = \frac{(2\pi)^3 T^2_{\rm sys}}{\Delta f t_{\rm obs} f^2_{\rm cover} \ell_{\rm max}(\nu)^2} \, ,
\end{equation} where $T_{\rm sys}$ is the system temperature,
$\Delta f$ is the chosen frequency (equivalently, redshift) window, and $t_{\rm obs}$ is the total observation time;  $\ell_{\rm max}(\nu)$ is the  highest multipole that can be measured by the array at frequency $\nu$ (wavelength $\lambda$), and is related to $D_{\rm tel}$, the maximum baseline of the core array, by $\ell_{\rm max}(\lambda)=2\pi D_{\rm tel}/\lambda$.
 $f_{\rm cover}$ is the total collecting area of the core array, $A_{\rm coll}$ divided by $\pi(D_{\rm tel}/2)^2$. 

For the foreground lenses ($f$) we will consider a SKA1-Mid-like configuration (i.e. SKA-Mid Phase 1). For the background sources ($b_1,b_2$) we consider the LSST-like survey that was described in the previous Section.  We split the foreground population in bins of $\Delta f = 40 \, {\rm MHz}$, covering the redshift range $0.1<z<1$. The rest of the parameters are $A_{\rm coll} = 0.08 \, {\rm km}^2$, $D_{\rm tel} = 8 \, {\rm km}$, $T_{\rm sys} = 20 + 66 \times \left(\frac{\nu}{300 \,{\rm MHz}}\right)^{-2.55} \, {\rm K}$, and $t_{\rm obs} = 10^4 \, {\rm hrs}$ \cite{Dewdney13}. The observed IM auto power spectrum is given by
\be
C^{\rm obs}_{\delta_{\rm IM} \delta_{\rm IM}}(\ell,f)=[\bar{T}(z_f)]^2 b^2_{\rm HI} C_{\delta \delta}(\ell,f)+C^{\rm N}_\ell,
\ee where we used the fact that the redshift slices are thin and the mean brightness temperature 
\be
\bar{T}(z)=566 \left(\frac{\Omega_{\rm HI}(z)h}{0.003}\right)\frac{(1+z)^2}{H(z)/H_0} \, {\rm \mu K}
\ee varies slowly within each bin.  
Therefore $N_f(\ell)$ is given by (setting $b_{\rm HI}=1$)
\be
N_f(\ell) = \frac{C^{\rm N}_\ell }{[\bar{T}(z_f)]^2}.
\ee 

We first compare the performance of combining SKA1-Mid (for the foreground lenses) and LSST (for the background sources) with the performance of the LSST-only case presented in the last subsection. 
In order to model the various power spectra we use the non-linear fitting formula by Smith et al. \cite{Smith:2002dz} and set the HI bias $b_{\rm HI}=1$; we also assume a constant $\Omega_{\rm HI}=4.9 \times 10^{-4}$ which is the value calculated from the local HI mass function measured by HIPASS \cite{Zwaan:2003hp}. 
For the same assumptions of the LSST analysis (in particular $\ell_{\rm max}=3000$), replacing the LSST low redshift density measurement with SKA1-Mid observations provides constraints very consistent with those seen in 
Figure~\ref{fig:LSSTconstr}.

\begin{figure}[H]
\centering
\includegraphics[scale=0.6]{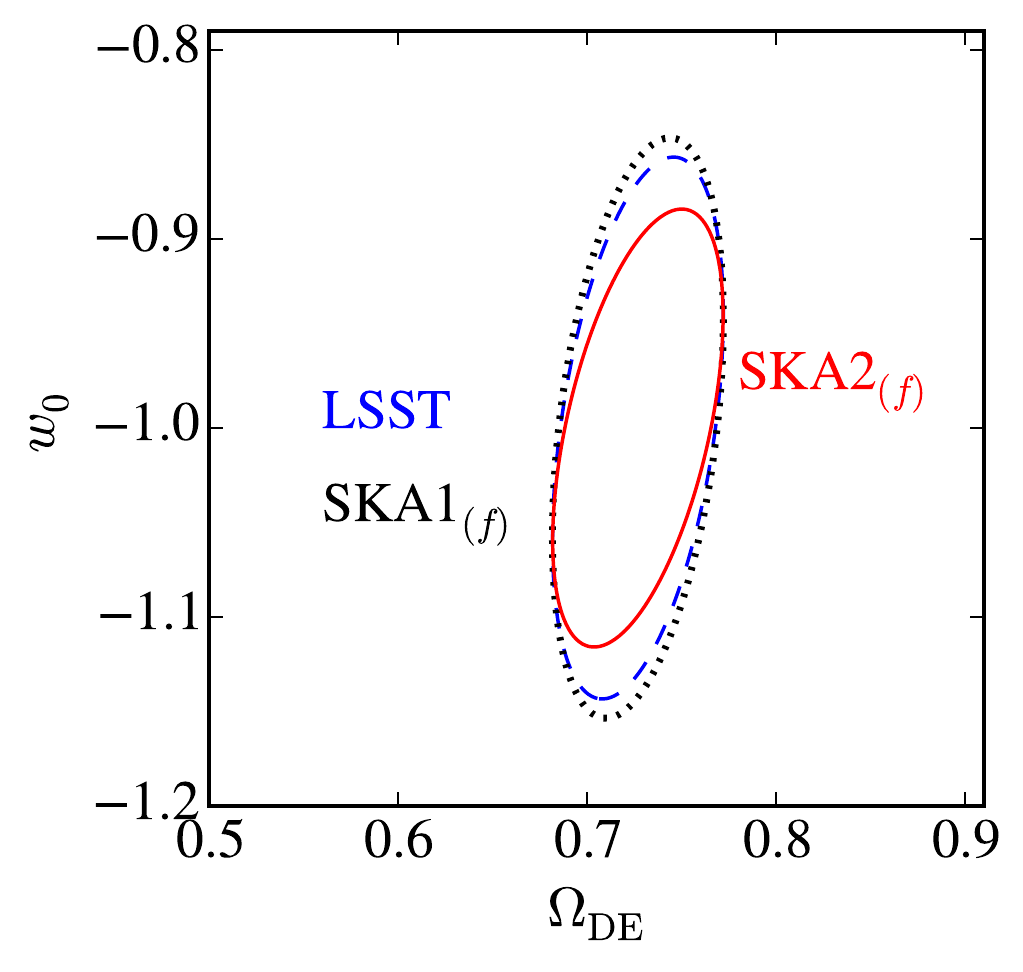}
\caption{Dark energy $1\sigma$ constraints on $w_0$ and $\Omega_{\rm DE}$ using the same assumptions as in Fig.~\ref{fig:LSSTconstr}. We show a comparison of the LSST-only results (dashed blue contour, labelled ``LSST") with the results of combining SKA1-Mid for the foreground lenses and LSST for the background sources (dotted black contour, labelled ``SKA1$_{(f)}$") using $\ell_{\rm max}=10^4$. Results using SKA2-Mid for the foreground lenses are also shown (solid red contour, labelled ``SKA2$_{(f)}$"). }
\label{fig:SKA2_Mid_constr_f}
\end{figure}

In order to utilise the smaller angular scales probed we next push the analysis to a larger value of $\ell_{\rm max}$ to $10^4$. As we can see in Fig.~\ref{fig:SKA2_Mid_constr_f}, the results are improved. We also present results considering the SKA2-Mid case (i.e. $10$ times bigger collecting area than SKA1) and $\ell_{\rm max}=10^4$. At very high $\ell$ values the shape noise terms dominate, and taking $\ell_{\rm max} > 10^4$ does not significantly improve the constraints. We should also note that since we use power spectra modelling to calculate the variance terms, the conservative approach would be to use a smaller $\ell_{\rm max}$ to stay away from non-linear scales. However, we feel this is too conservative as the ratio $R$ is $\ell$-independent and the power spectra that enter the error calculation are known with quite good accuracy that will improve considerably with future observations.

 The $1\sigma$ marginalised results for the constant equation of state parameter uncertainty are
\begin{align} \nonumber
&\delta w_0|^{\rm SKA1_{(f)}}_{\ell_{\rm max}=3000} \simeq \delta w_0|^{\rm LSST}_{\ell_{\rm max}=3000} \simeq 0.12, \\ \nonumber
&\delta w_0|^{\rm SKA1_{(f)}}_{\ell_{\rm max}=10^4} \simeq \delta w_0|^{\rm LSST}_{\ell_{\rm max}=10^4}  \simeq 0.10, \\
&\delta w_0|^{\rm SKA2_{(f)}}_{\ell_{\rm max}=10^4} \simeq 0.08.
\end{align}
Here we note that our results for the LSST-only case with $\ell_{\rm max}=10^4$ are in agreement with previously published geometrical constraints, although an exact comparison is difficult to make as different works have used different methods, assumptions, approximations and priors. Our approach is closer to the one followed by \cite{Zhang:2003ii}. There, a constraint $\delta w_0 \sim 0.1$ is found for a survey with parameters similar to LSST.

Using SKA1 for the foreground measurement offers comparable constraints to using LSST, but the cross-correlations could be less susceptible to systematic contamination; improving the observations assuming SKA2 parameters can improve the constraints. However, the improvement is not very large as the dominant terms in the Fisher matrix calculations are the LSST shape noise terms shown in Fig.~\ref{fig:CgCk}, bottom panel.

\subsection{Results using a LSST-like survey for ($f, b_1$) and a SKA-Mid-like Intensity Mapping survey for ($b_2$)}

We next consider the case where we probe the foreground ($f$) lenses and the $(b_1)$ sources with a galaxy survey like LSST, but we use a SKA-Mid-like Intensity Mapping survey for the ($b_2$) sources. In order to calculate the noise in the $\kappa$ measurement for this case, we use the quadratic estimator technique developed in \cite{Pourtsidou:2013hea, Pourtsidou:2014pra}. This estimator takes into account the discreteness and clustering of galaxies and is applicable to any redshift below reionization. Using the intensity mapping approach and an SKA-like array, the method performs better than 21cm galaxy surveys at high redshifts. Hence, in the following we are going to apply the weak lensing intensity mapping estimator for the second background sources $(b_2)$ at various redshifts $z>2$, but still use LSST for the closer background sources ($b_1$) at $z \sim 1$. We are also going to use LSST for the foreground lenses ($f$), but keeping in mind that an SKA-like interferometer can perform equally well or better, as shown in Fig.~\ref{fig:SKA2_Mid_constr_f}.

For a detailed discussion and derivation of the results used here, see \cite{Pourtsidou:2014pra}. The expression for the lensing reconstruction noise $N_{b_2}(\ell)$ from the intensity mapping method is rather lengthy, so we will not include it here, but the interested reader is referred to \cite{Pourtsidou:2014pra}, Appendix C. To summarise, $N_{b_2}$ involves the underlying dark matter power spectrum $P_{\delta \delta}$, the HI density $\Omega_{\rm HI}(z)$ as well as the HI mass (or luminosity) moments up to 4th order and, of course, the thermal noise of the instrument (in our case the SKA).
In \cite{Pourtsidou:2014pra} it was found that the signal-to-noise is strongly dependent on the possible evolution of the HI mass function. More specifically, assuming the no-evolution scenario (which is the most conservative, but also less realistic approach), precise measurements can be made with an SKA2-like instrument; however if we assume instead a model where the HI density $\Omega_{\rm HI}(z)$ increases by a factor of $5$ by redshift $z=3$ and then slowly decreases towards redshift $z=5$, as suggested by the DLA observations from \cite{Peroux:2001ca}, high signal-to-noise can be achieved even with SKA1.

\begin{figure}[H]
\centering
\includegraphics[scale=0.65]{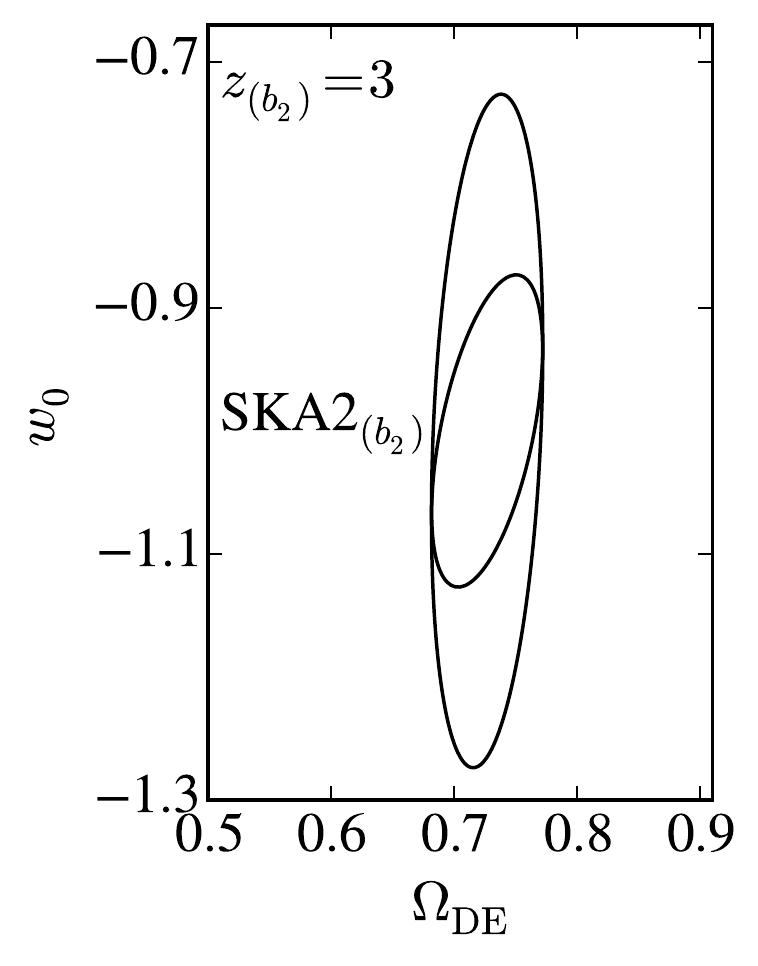}
\caption{Dark energy $1\sigma$ constraints on $w_0$ and $\Omega_{\rm DE}$ Dark energy $1\sigma$ constraints on $w_0$ and $\Omega_{\rm DE}$ using the same assumptions as in Fig.~\ref{fig:LSSTconstr}. We have used a combination of a LSST-like galaxy survey in the foreground and the $b_1$ background, and a SKA2-Mid-like IM survey in the $b_2$ background with parameters described in the main text and $\ell_{\rm max}=10^4$. Sources are at $z_{b_2}=3$. We show results assuming no HI evolution (larger contour) and using the evolution scenario described in the main text (smaller contour).}
\label{fig:SKA2_Mid_constr_b2}
\end{figure}

The parameters for SKA2-Mid are the same as in the previous subsection. The source redshift is $z_{b_2}=3$ with a bandwidth of $40 \, {\rm MHz}$. The results are shown in Figure~\ref{fig:SKA2_Mid_constr_b2}, and we have used $\ell_{\rm max}=10^4$. The largest contour corresponds to the no-evolution HI scenario, while the smaller one assumes the aforementioned HI evolution model. 

The results are again competitive with the LSST-only case, and they considerably benefit from a possible HI evolution. The result for HI sources at $z=3$ using the evolution model and SKA2-Mid is $\delta w_0 \simeq 0.08$. 

\subsection{Results using a LSST-like survey for ($f, b_1$) and 21cm lensing from the Epoch of Reionization ($b_2$)}

Let us now consider the case where our second background ($b_2$) sources are at the Epoch of Reionization (EoR). This provides a longer lever arm for the ratio $R$ and uses information from an era that will be explored for the first time with the SKA. 
The possibility of measuring the lensing signal with 21cm emission from the EoR has been studied previously. In \cite{Zahn:2005ap} and \cite{Metcalf:2008gq} the  convergence
estimator and the corresponding lensing reconstruction noise were calculated assuming that the temperature (brightness) distribution is Gaussian, which is a reasonable approximation at the EoR, at least while the ionised regions are small. 
An important advantage of 21cm lensing is that one is able to combine information from multiple redshift slices. 

In Fourier space, the temperature fluctuations are divided into wave vectors perpendicular to the line of sight $\mathbf{\kpe}=\mathbf{l}/{\cal D}$, with ${\cal D}$ the angular diameter distance to the source redshift, and a
discretised version of the parallel wave vector $\kpa =\frac{2\pi}{{\cal L}}j$ where ${\cal L}$ is the depth of the observed volume. Considering modes with different $j$ as independent, an optimal
estimator can be found by combining the individual estimators for
different $j$ modes without mixing them. The three-dimensional lensing reconstruction noise is then found to be \citep{Zahn:2005ap}
\begin{align} \nonumber
&N_{b_2}(\ell) = (\ell^4/4) \times \\
&\left[\sum_{j=1}^{j_{\rm max}} \int \frac{d^2\ell'}{(2\pi)^2}  \frac{[\bl' \cdot \bl C_{\ell',j}+\bl \cdot (\bl-\bl')
C_{|\ell'-\ell|,j}]^2}{2 C^{\rm tot}_{\ell',j}C^{\rm tot}_{|\ell'-\ell|,j}}\right]^{-1},
\end{align}
where
\be
\label{eq:Cellj}
C_{\ell,j} = [\bar{T}(z)]^2 \frac{P_{\delta \delta}(\sqrt{(\ell/{\cal D})^2+(j2\pi/{\cal L})^2})}{{\cal D}^2 {\cal L}}
\ee  and
\be
C^{\rm tot}_{\ell, j} = C_{\ell,j} + C^{\rm N}_\ell.
\ee

\begin{figure}[H]
\centering
\includegraphics[scale=0.65]{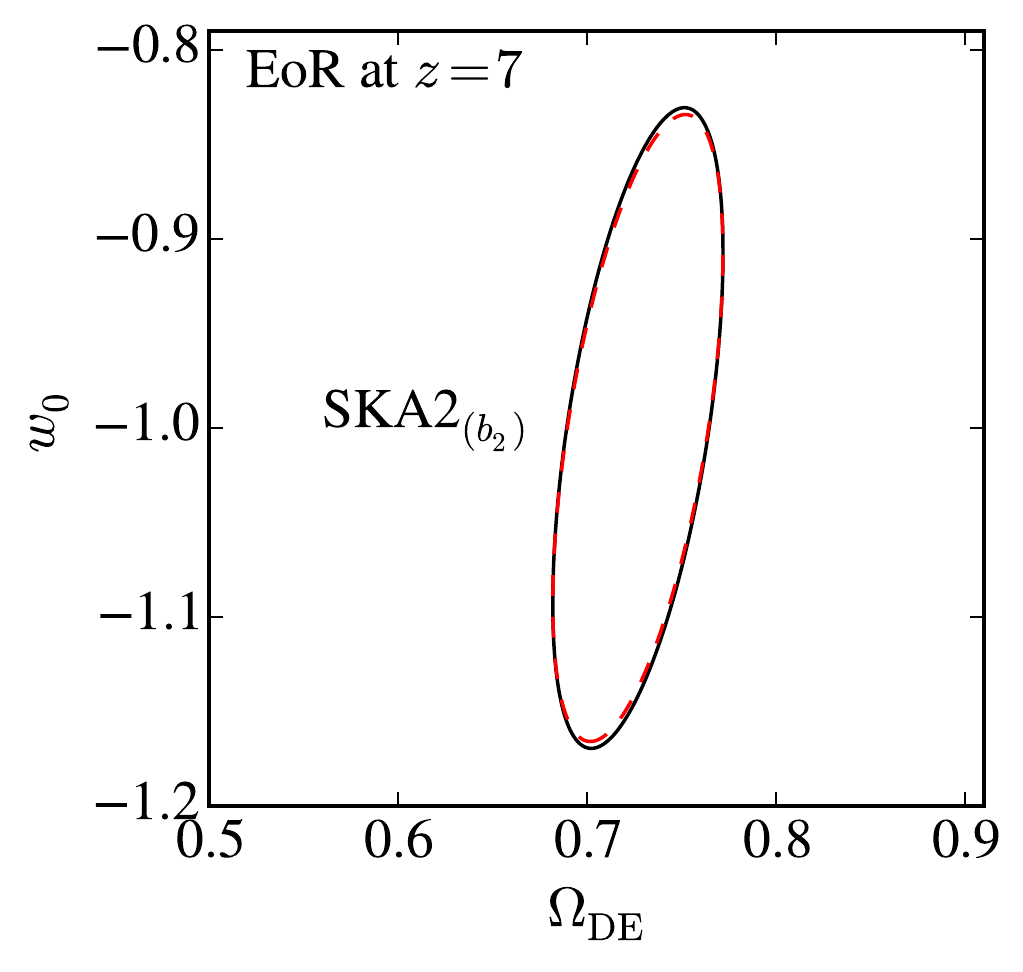}
\caption{Dark energy $1\sigma$ constraints on $w_0$ and $\Omega_{\rm DE}$ using the same assumptions as in Fig.~\ref{fig:LSSTconstr}. We have used a combination of a LSST-like galaxy survey in the foreground and the $b_1$ background, and a SKA2-Mid-like instrument in the $b_2$ background at the EoR redshift assumed to be $z=7$ (solid black contour). We also show results with a more compact SKA-like instrument (dashed red contour).}
\label{fig:SKAlike_constr_b2_EoR}
\end{figure}

Following this approach, we consider the case where we probe the foreground ($f$) lenses and the $(b_1)$ sources with a galaxy survey like LSST, and we use a SKA-like instrument for the ($b_2$) sources at the EoR redshift, which we assume to be instantaneous at $z_{b_2}=7$.
In order to facilitate the comparison with the previous Sections, we are initially going to consider a SKA2-Mid-like instrument but with a smaller observation bandwidth equal to $8 \, {\rm MHz}$. That is because the Gaussian quadratic lensing estimator is optimised for the case where the statistical properties of the 21cm radiation signal and noise are constant within a band, and an observation bandwidth of a few MHz is small enough so that this assumption is justified. The results are shown in Figure~\ref{fig:SKAlike_constr_b2_EoR} for $\ell_{\rm max}=10^4$ (solid black contour). They give a marginalised uncertainty $\delta w_0 \simeq 0.11$.

However, SKA-Mid is not designed to probe low enough frequencies (equivalently, high redshifts), and the current plans for SKA-Low which will observe the EoR redshifts will scan a very small sky area hence it will not be able to achieve the required precision in the $\kappa$ measurement. 
In \cite{Pourtsidou:2013hea, Pourtsidou:2014pra} it was found that a compact SKA-like instrument, i.e. an array with smaller maximum baseline and a smaller core collecting area, would perform a weak lensing intensity mapping survey equally well or even better than the current SKA-Mid design. This means that the precision required could be achieved by using a more compact configuration. 
We therefore present results considering a SKA-like ``purpose-built" compact interferometer. 
The collecting area is $A_{\rm coll}=0.5 \, {\rm km}^2$ and the maximum baseline $D_{\rm tel}=4 \, {\rm km}$, keeping the rest of the parameters the same. The results are shown in Figure~\ref{fig:SKAlike_constr_b2_EoR} (dashed red contour).  
With a more compact configuration, we can obtain identical constraints from cross-correlation cosmography with significantly less collecting area ($A_{\rm coll}=0.5 \, {\rm km}^2$, compared to $A_{\rm coll}=0.8 \, {\rm km}^2$ assumed in the original SKA2 configuration.)  

\subsection{Results using a LSST-like survey for ($f, b_1$) and CMB lensing with a COrE-like satellite for ($b_2$)}

\begin{figure}[H]
\centering
\includegraphics[scale=0.7]{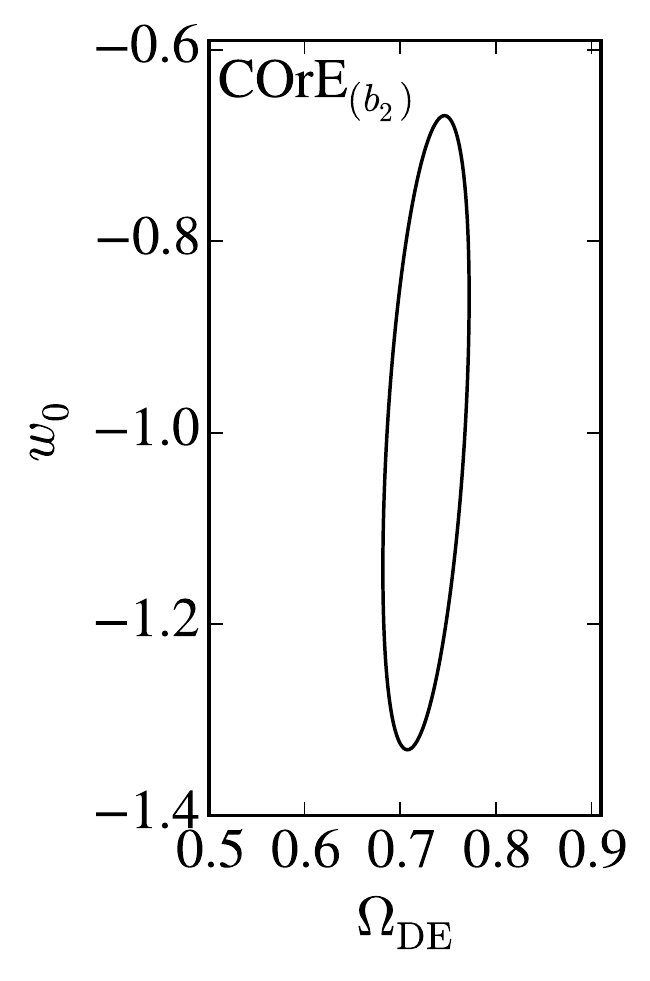}
\caption{Dark energy $1\sigma$ constraints on $w_0$ and $\Omega_{\rm DE}$ using the same assumptions as in Fig.~\ref{fig:LSSTconstr}.  We have used a combination of a LSST-like galaxy survey in the foreground and the $b_1$ background, and a COrE-like CMB survey in the $b_2$ background.}
\label{fig:constr_b2_cmb}
\end{figure}

Finally, we consider the case where we use the CMB as the second source plane. This possibility has been studied in the past \cite{Hu:2007jh,Das:2008am} and its main advantages are the very well determined source redshift and distance and the fact that it provides the largest possible lever arm for the distance ratio. For the CMB case, the two-dimensional lensing reconstruction noise is found to be \cite{Hu:2001kj}
\begin{align}\nonumber
&N(\ell)|_{\rm cmb} = (\ell^4/4) \times \\
&\left[\int \frac{d^2\ell'}{(2\pi)^2}  \frac{[\bl' \cdot \bl C^{\rm TT}_{\ell'}+\bl \cdot (\bl-\bl')
C^{\rm TT}_{|\ell'-\ell|}]^2}{2 C^{\rm tot}_{\ell'}C^{\rm tot}_{|\ell'-\ell|}}\right]^{-1},
\end{align} where $C^{\rm TT}_{\ell}$ is the CMB TT power spectrum and 
\be
C^{\rm tot}_{\ell} = C^{\rm TT}_{\ell} + C^{\rm N}_\ell,
\ee with $C^{\rm N}_\ell$ the instrumental noise for the CMB survey. We will consider a future COrE-like satellite with FWHM $\sigma = 3.0^\prime$ and temperature noise $\Delta_T = 1 {\, \rm \mu K^\prime}$. Then we have
\be
 C^{\rm N}_\ell = \Delta^2_T \, {\rm exp}[\ell(\ell+1)\sigma^2/8{\rm ln}2].
\ee 
In Figure~\ref{fig:constr_b2_cmb} we show the derived constraints using the aforementioned satellite for the second source plane ($b_2$) and the LSST-like survey for ($f, b_1$). The marginalised $w_0$ uncertainty is found to be $\delta w_0 \simeq 0.22$. Comparing these results to the SKA-like EoR case from the previous Section we see that the latter give much better constraints. That is because the CMB reconstruction comes with higher statistical errors compared to the 21cm reconstruction, and the leverage we get from the higher redshift is not enough to compensate. 

\subsection{Noise terms comparison}

\begin{figure}[H]
\centering
\includegraphics[scale=0.6]{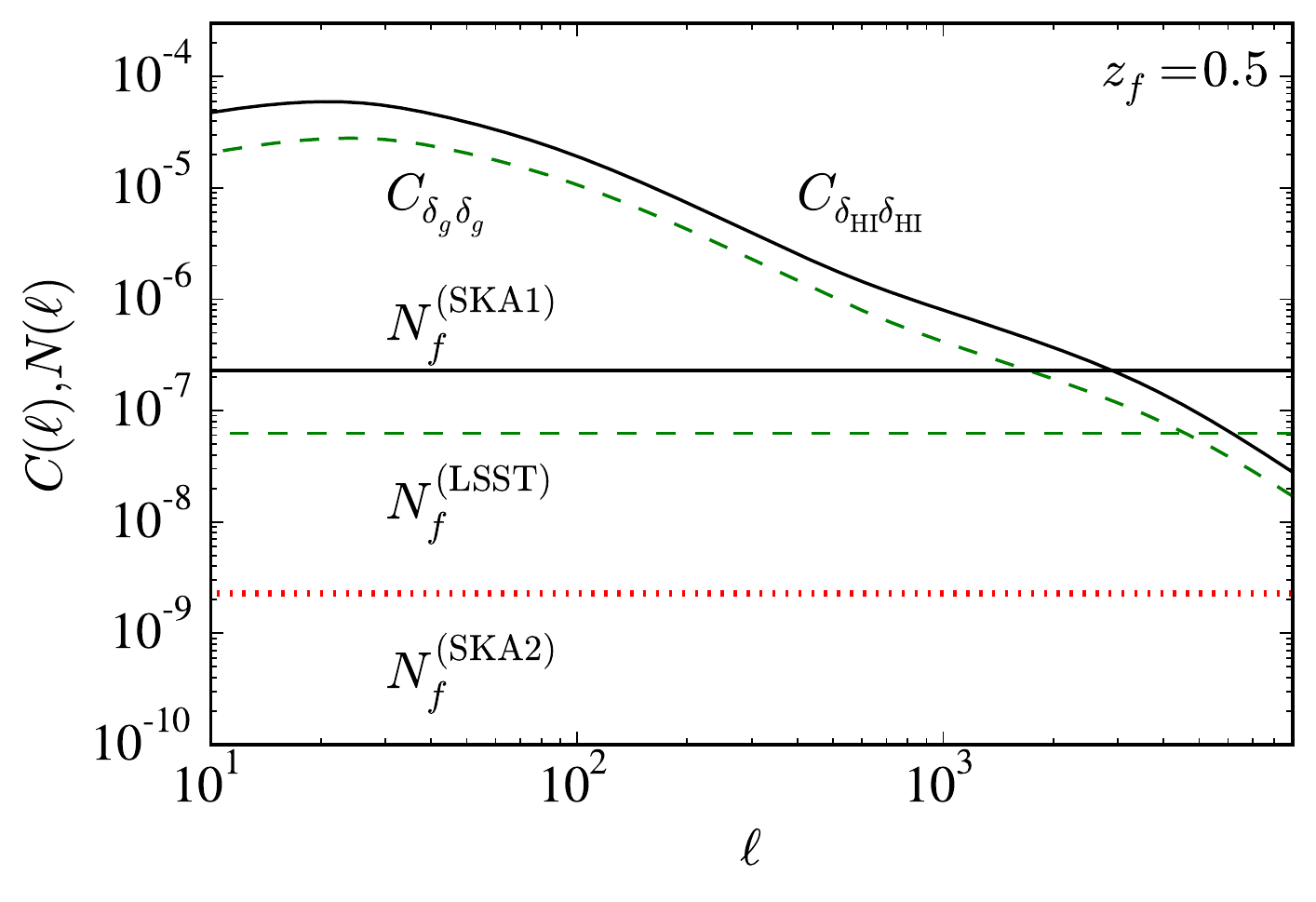}
\includegraphics[scale=0.6]{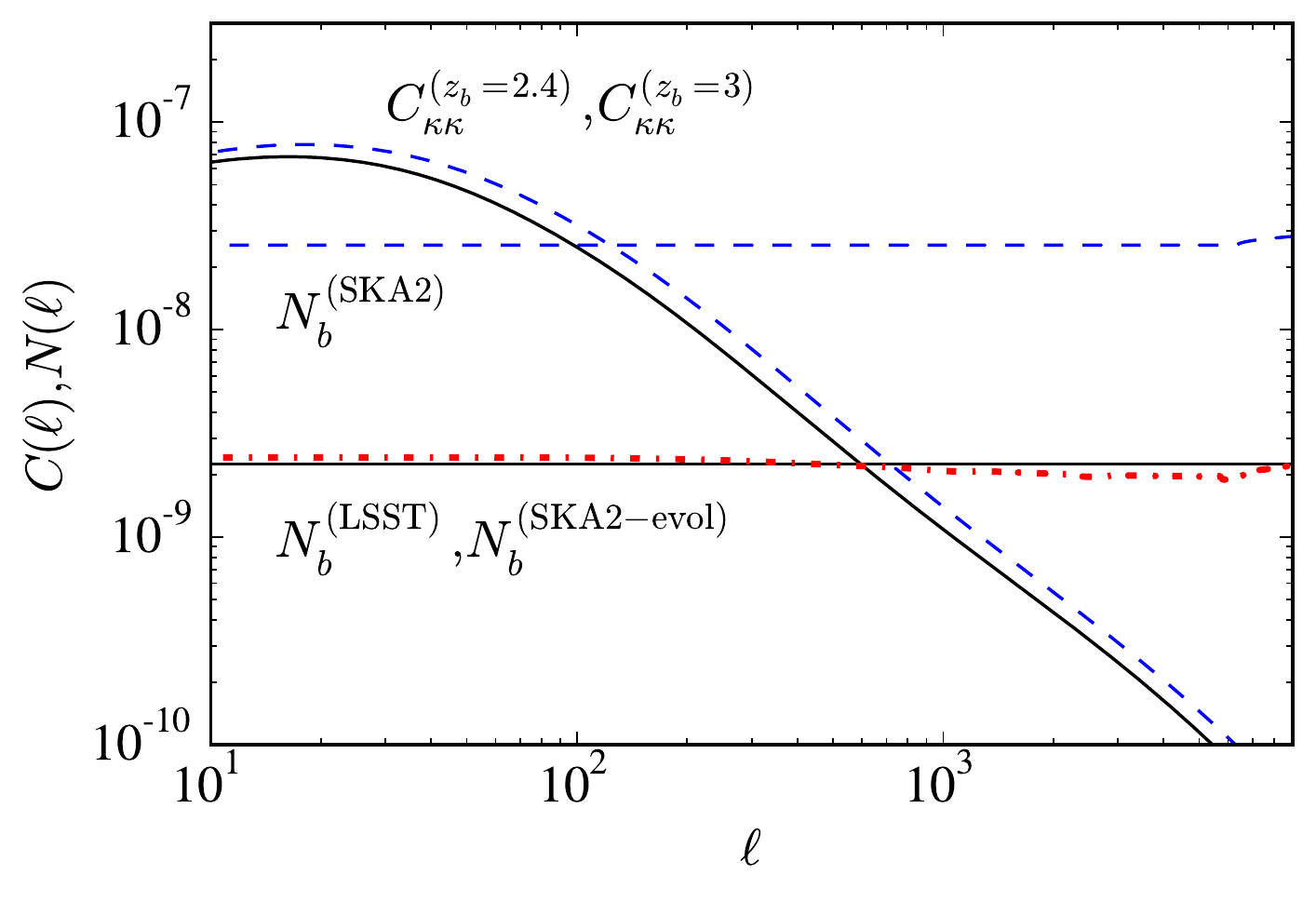}
\includegraphics[scale=0.6]{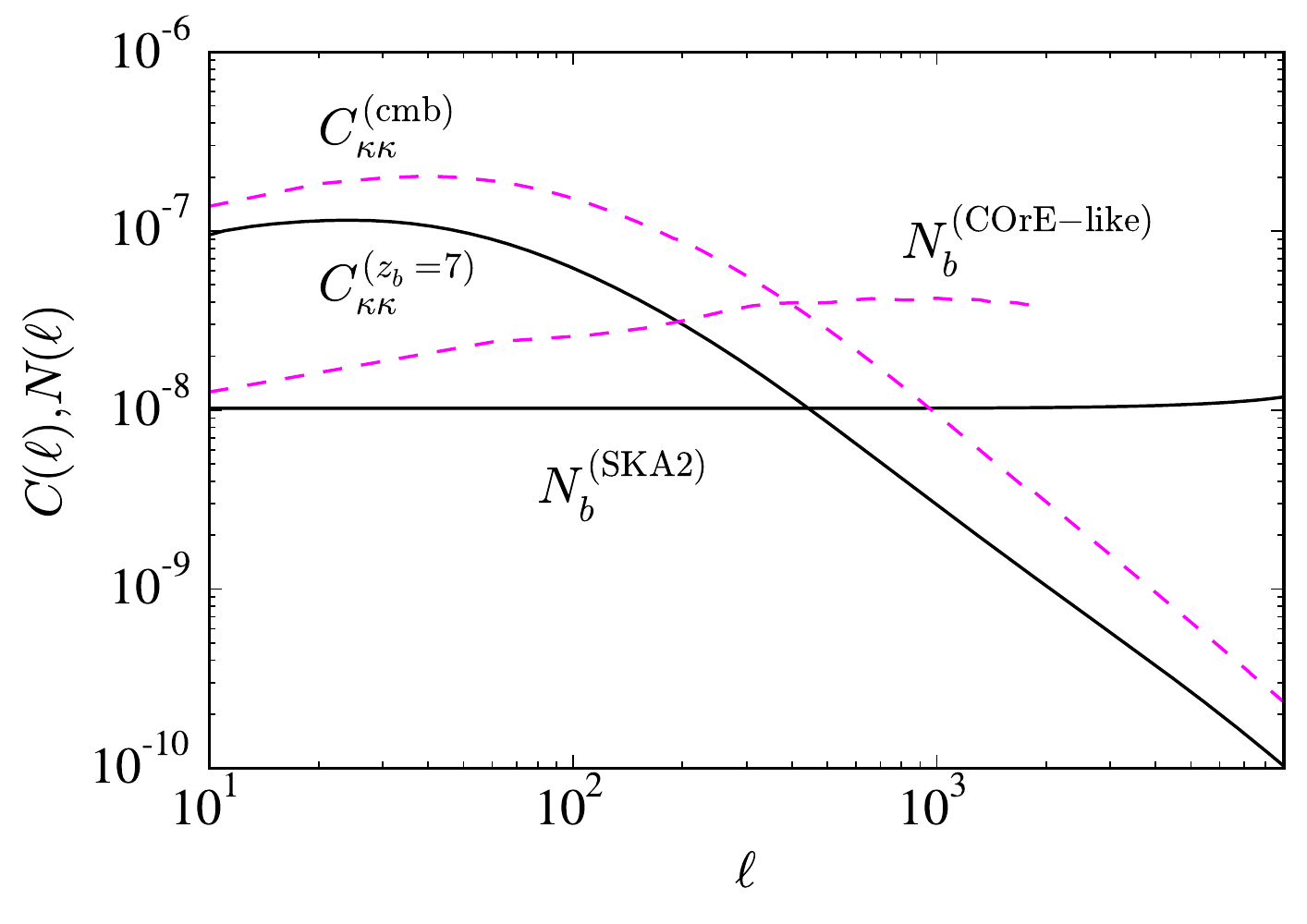}
\caption{A comparison of the noise terms for the various survey combinations considered. See text for further details.}
\label{fig:noisecomp}
\end{figure}

To consolidate our results, we compare the noise terms used in the Fisher matrix calculations for the various survey combinations we have considered. The top panel of Fig.~\ref{fig:noisecomp} compares the tracer density power spectra and noise terms for LSST (dashed green lines), SKA1 (solid black lines) and SKA2 (dotted red line) at the $z_f = 0.5$ foreground slice. $C_{\delta_g \delta_g}$ and $C_{\delta_{\rm HI} \delta_{\rm HI}}$ are comparable but not exactly the same because their redshift binnings differ slightly;  for LSST, $\Delta z=0.1$ but the choice of $\Delta f =40 {\rm MHz}$ bandwidth for the intensity mapping survey corresponds to a narrower bin at $z_f=0.5$. The LSST and SKA1 noise terms ($N_f$) are comparable, while the SKA2 noise is much lower. This explains the results of Fig.~\ref{fig:SKA2_Mid_constr_f}, where we found comparable constraints from LSST-only and $\rm{SKA1}_{(f)}$ cases, and stronger constraints using $\rm{SKA2}$ to map the foreground density.

For the intermediate lensing data, which we typically assume is centred at $z_{b_1}=1.3$ to maximise the distance to the higher redshift lensing, the HI lensing noise  is not competitive with that of the LSST.   However, the HI intensity mapping measurements can be competitive for highest redshift sample, where they can observe at higher redshifts than are possible with LSST.  The middle panel of Fig.~\ref{fig:noisecomp} compares the convergence power spectra and $N_{b_2}$ noise terms for LSST at $z_{b_2}=2.4$ (solid black lines), SKA2 at $z_{b_2}=3$ assuming no HI evolution (dashed blue lines) and SKA2 at $z_{b_2}=3$ assuming the HI evolution model described in the main text (dotted-dashed red line). As we can see, the LSST and SKA2 HI evolution noise terms are comparable, and in the SKA2 case we also have a longer lever arm because the $b_2$ sources are at higher redshift.

Finally, the bottom panel of Fig.~\ref{fig:noisecomp} compares the convergence power spectra and $N_{b_2}$ noise terms for a SKA2-like instrument at the EoR redshift assumed to be $z=7$ (solid black lines) and a COrE-like satellite with sources at the CMB redshift $z=1090$ (dashed magenta lines). We see that the noise level of the COrE-like mission is higher than SKA2 and quickly diverges as we reach the limits ($\ell_{\rm max} \simeq 3600$), set by the beam resolution.  Again, this is a consequence of the CMB providing only a single source plane, while the HI data offers multiple source planes that can be co-added to reduce the noise.

\section{Conclusions}
\label{sec:conclusions}

In this paper, we have shown how the HI intensity mapping technique can be used to enhance cross-correlation cosmography studies. Cross-correlation cosmography is based on the scaling of the cross-correlation signal 
$C_{\delta_{\rm tr} \kappa}$ with the redshift of the background source population. The ratio of the signals from the same foreground lens population to two different background populations simplifies to a geometrical distance ratio that only depends on dark energy parameters and curvature. Intensity mapping is a technique that treats the 21cm emission as a continuous unresolved background, without resolving or even identifying (in angular resolution, not frequency) individual galaxies. It offers excellent redshift resolution and a longer lever arm for the background sources. 

We performed a comprehensive study of the possibility of combining optical galaxy surveys, in particular LSST, with HI intensity mapping surveys using a SKA-like instrument in order to derive geometrical dark energy constraints. Our results show that using the SKA to measure the foreground density field $f$ and/or the background convergence field $b_2$ at high redshifts $z>2$ can be beneficial. More specifically, we find that a constant equation of state for dark energy can be constrained to $\sim 8\%$ for a sky coverage $f_{\rm sky} = 0.5$ and assuming a $\sigma (\Omega_{\rm DE})=0.03$ prior for the dark energy density parameter. One major uncertainty is the unknown evolution of the HI density parameter and the form of the HI mass function that is crucial for the modelling of the lensing reconstruction noise using intensity mapping. However, the no-evolution model we considered is the most conservative scenario. Using a more optimistic - but also more realistic - evolution scenario significantly improves the expected constraints.  The Epoch of Reionization itself can potentially be used as our second background source; it provides a longer lever arm for the distance ratio and the combination of the contribution from many redshifts slices results in a low noise level, offering significantly better constraints compared to using the CMB, even when assuming COrE-like satellite observations.  

Looking at Table~\ref{tab:resultssumm} which summarizes all the survey combinations we have considered, one notices that we never consider all ($f,b_1,b_2$) populations to be probed with intensity mapping; we always use an LSST-like survey for the closer background population $b_1$. That is because the weak lensing intensity mapping estimator is competitive with optical galaxy surveys at high redshifts $z>2$, where we also have the advantage of a longer lever arm. In lower redshifts a galaxy survey performs better. We also note that in cases III.C, III.D and III.E in Table~\ref{tab:resultssumm} we can replace LSST with SKA1 for the foreground lenses ($f$), as they perform equally well (see Fig.~\ref{fig:SKA2_Mid_constr_f}).

Foreground contamination can be a significant concern. Foregrounds are the most important source of systematic error when it comes to the intensity mapping technique, but interferometric techniques can filter them out. In particular for our lensing studies, where the frequency-dependent foreground contribution is large, it has been shown that the foreground subtraction techniques will remove parallel $k$ modes \cite{McQuinn:2005hk,Zahn:2005ap}, meaning that the lensing reconstruction noise will increase somewhat. However, if the subtraction techniques are successful, only the first few modes will be removed and the signal-to-noise of the measurements will not be significantly affected \cite{Pourtsidou:2014pra}. 

To conclude, we have demonstrated that HI intensity mapping can be used for cosmographic studies, to complement and compete with the state-of-the art optical galaxy surveys. In addition, the weak lensing intensity mapping technique performs very well across a wide range of post-reionization redshifts, and by using tomographic information in the measured convergence it is possible to infer how the matter power spectrum and the growth function evolve with time \cite{Pourtsidou:2014pra}. These two avenues (cosmography and lensing tomography) can be explored simultaneously in order to derive combined growth-geometry constraints on the dark energy parameters. 
An example of how this can be done is presented in \cite{Zhang:2003ii}. The dark energy parameters can be split into two kinds, those that enter the growth factor, and those that enter the geometrical distances. Such a splitting allows a robust consistency check if the equation of state $w$ values obtained separately from geometry and from growth are in agreement; if, however, they disagree, we can identify and remove important systematics or contaminations, or revise incorrect assumptions about the behaviour of the mass fluctuations. The results of such a consistency test are shown in \cite{Zhang:2003ii}. 

Our studies would benefit from the modelling of a ``realistic", non-instantaneous reionization history, from more accurate measurements of $\Omega_{\rm HI}b_{\rm HI}$ as a function of redshift and from detailed simulations of foreground contamination and subtraction techniques. These investigations are under way and will be presented in future work.

\section{Acknowledgments}
This work was supported by STFC grant ST/H002774/1. The authors would like to thank Robert Benton Metcalf for very useful discussions and comments.

\bibliographystyle{apsrev}
\bibliography{references_21cm}

\end{document}